\documentclass[twocolappendix,iop,apj]{emulateapj}
\usepackage{amsfonts,amsmath,graphicx,natbib,url,hyperref,nicefrac}
\hypersetup{colorlinks=true, urlcolor=blue, citecolor=blue}

\usepackage[usenames,dvipsnames]{color}
\usepackage{comment}

\newcommand{\gram}{\mbox{$\rm g$}}
\newcommand{\pcm}{\mbox{$\rm cm^{-1}$}}

\shorttitle{How Dense a CSM is Sufficient to Choke a Jet?} 
\shortauthors{Duffell \& Ho}

\begin{document}

\title{How Dense a CSM is Sufficient to Choke a Jet?}

\def\harv{1}
\def\cal{2}

\author{Paul C. Duffell\altaffilmark{\harv} and Anna Y. Q. Ho\altaffilmark{\cal}}
%\affil{Astronomy Department and Theoretical Astrophysics Center, University of California, Berkeley, CA 94720}
\altaffiltext{\harv}{Harvard-Smithsonian Center for Astrophysics, 60 Garden Street, Cambridge MA 02138}
\altaffiltext{\cal}{California Institute of Technology, 1200 E. California Boulevard, Pasadena CA 91125}
\email{paul.duffell@cfa.harvard.edu}

\begin{abstract}

The progenitor stars of stripped-envelope high-velocity supernovae (Ic-BL SNe) can explode inside 
a dense circumstellar medium (CSM) that extends out to many times the progenitor radius.
This complicates the question of
whether all Ic-BL SNe harbor a jet,
which can tunnel through the star and be viewed on-axis as a long-duration gamma-ray burst (GRB).
More specifically, a sufficiently dense CSM might ``choke'' the jet, redistributing its energy quasi-spherically.
In this study, we numerically calculate the CSM density necessary for jet-choking.  For typical GRBs, we determine the jet is not choked in the CSM unless $\rho r^2 > 4 \times 10^{19}\,\gram\,\pcm$; this requires several solar masses of CSM to be situated within $10^{13}$ cm of the progenitor, a much higher density than any CSM observed.  We conclude that typical GRB jets are not choked in the CSM.  However, in many cases the CSM has sufficient mass to decelerate the jet to a modest Lorentz factor ($\Gamma\sim10$), which should lead to a long coasting phase for the jet, observable as a long plateau (potentially up to a few days) in the afterglow light curve.  For extreme cases of low-energy GRBs in a high-mass CSM, the jet will decelerate to nonrelativistic velocities, causing it to spread modestly to a larger opening angle ($\theta_j \approx 20$ degrees) before breaking out of the CSM.  Even in these extreme examples, the jet does not have time to redistribute its energy quasi-spherically in the CSM before breakout.

\end{abstract}


\section{Introduction} \label{sec:intro}

In the traditional model of a core-collapse supernova, the core collapses and a shock unbinds the star with an energy release of $\sim10^{51}\,$erg.
A subset of CC SNe show no evidence for hydrogen and helium in their spectra; these are thought to arise from progenitors that have been stripped of their outer envelopes, and are classified as Type Ic \citep{Filippenko1997}.
A subset of Type Ic SNe are unusually fast and energetic: their photospheric velocities are a factor of two or more larger than typical CC SNe, and their kinetic energies an order of magnitude greater.
Their spectral features are also broader than that of typical Ic spectra \citep{Modjaz2016} giving rise to the classification ``broad-lined" Ic (Ic-BL).

The reason for these large energies and high velocities is unknown,
but one clue lies in the fact that Ic-BL SNe are the only type of SN ever observed in conjunction with a long-duration gamma-ray burst (GRB; \citealt{WoosleyBloom,2017AdAst2017E...5C}).
In fact, it has been suggested that all Ic-BL SNe harbor a jet, and that a GRB is the unusual case in which the jet successfully tunnels through the star and is viewed on-axis \citep{Sobacchi2017,2018ApJ...860...38B}.

Observations of some Ic-BL SNe have shown evidence for high densities in the immediate vicinity of the progenitor star.
One line of evidence is double-peaked optical light curves \citep{Piro2015},
with a first peak from interaction and the second peak from the radioactive decay of Ni-56.
This was first observed in SN2006aj \citep{2006Natur.442.1008C}, which was
inferred to have a CSM mass of
$0.01 M_{\sun}$ distributed within $\sim 3 \times 10^{13}$ cm \citep{2014ApJ...788..193N}.
A larger CSM mass can extend the first peak,
blending it into the second peak---
the result is a light curve too rapidly rising and luminous to be explained by the usual radioactive decay.
This was first seen in iPTF16asu \citep{2017ApJ...851..107W} and then again in SN2018gep \citep{2019arXiv190411009H}.

Another line of evidence is luminous radio and X-ray emission, as in PTF11qcj \citep{2014ApJ...782...42C}, SN2007bg \citep{2013MNRAS.428.1207S}, and SN2003L \citep{2006ApJ...651.1005S}.  Analysis of GRB afterglow spectra has also pointed towards high-density CSM surrounding their progenitors \citep{2015ApJ...805..159M}.

Given that GRBs and Ic-BL SNe have been observed to have high-density ambient media,
it is worth asking how this environment might alter the dynamics of a GRB jet.  In particular, \cite{2015ApJ...807..172N} posited that a GRB jet would have to be choked in such a dense environment,
and that these ``choked jets'' could explain the class of low-luminosity GRBs that have been observed at low redshift \citep{2015ApJ...807..172N}
and which could be a source of high-energy neutrinos \citep{Meszaros2001}.

In this paper, we define ``choked'' as follows.
As the jet tunnels through the star, the ram pressure in the jet balances with the thermal pressure in the cocoon. When the engine shuts off, the ram pressure is suddenly removed. The high-pressure cocoon then crushes the jet core, contaminating it with baryons and quickly thermalizing the jet kinetic energy. The flow from that point on mostly resembles a non-relativistic blastwave, which rapidly evolves toward a spherical explosion.  Jet choking is typically seen in calculations of a jet drilling through a star when the engine is shut off before breakout from the stellar surface \citep{2001ApJ...550..410M}.  Here we instead explore the idea of jet choking outside the stellar surface, in the dense CSM.

This idea that jets could be choked in the CSM has been applied to other observations.  After observations of a jet cocoon were reported associated with SN2017iuk \citep{2019Natur.565..324I}, \cite{2019Natur.565..300N} argued that the associated jet would have been choked in the extended CSM.  \cite{2019arXiv190502226T} speculated that a jet may have been choked in the dense CSM surrounding the energetic SN2016coi.

\cite{2015ApJ...807..172N} regarded the dense CSM around SN2006aj as an extension of the stellar envelope, meaning that the jet has a much longer distance to propagate before breakout than in normal (stripped) GRB progenitors.  In fact, interpreting the CSM in this way almost guarantees that any jet will be choked in this medium, unless the engine is left on for the time it takes to break out --- at least a light crossing time, which is 

\begin{equation}
    t_{\rm engine} > R_{\rm CSM}/c \sim 10^3 ~\rm s.
\end{equation}

Given that typical long-GRB durations are $\sim 10$ seconds, it stands to reason that all GRB jets would be choked in such a medium.
However, this cannot be the only criterion for jet choking, as clearly the CSM needs to be above some critical density to be capable of choking the jet. For example, the GRB afterglow jets we observe are in no danger of being ``choked" by the ISM in this fashion, even after the engine has long been shut off.

This study aims to calculate the critical density necessary for jet choking in the long GRB context.  In section \ref{sec:anly} we argue that this criterion should only depend on the ratio $L_{\rm iso}/A$, where $L_{\rm iso}$ is the isotropic equivalent engine luminosity, and $A = \rho r^2$ is the density parameter assuming the CSM is distributed in a wind.  In section \ref{sec:numerics} we calculate the critical value of $L_{\rm iso}/A$ using numerical calculations of a jet breaking out of a star in different CSM environments.  Additionally, we calculate how much the CSM decelerates the jet, and whether this deceleration is by itself sufficient to spread the jet significantly.  Finally, in section \ref{sec:disc} these results are synthesized and applied to interpretations of observed CSM around supernovae.

\section{Analytical Considerations}
\label{sec:anly}

In order for a jet to be choked when the engine is turned off, there must be some criterion on the local density.  For example, GRB afterglows can persist long after the engine is turned off.  In this case, the jet is not being actively collimated by the medium; instead of a ``cocoon" there is really just a bow shock, which does not take an active role in collimating the jet.  In this case, the jet will only decelerate and spread after it sweeps up sufficient mass.

This situation is distinct from when the engine is still pushing a jet through the progenitor star, because in that case the jet is being collimated by the star.  Only after breakout does collimation cease and the opening angle of the jet is given by the opening angle of the engine ($\theta_j = \theta_0$).  Therefore, the question of whether the engine turning off will affect the jet should be equivalent to the question of whether the jet is being collimated by the surrounding medium.

This question can be rephrased in the language of \cite{2011ApJ...740..100B}, which describes various regimes of jet propagation depending on the engine power and ambient density.  The relevant condition for a collimated vs. uncollimated jet is

\begin{equation}
    \tilde L \gtrsim \theta_0^{-4/3},
\end{equation}

where $\tilde L = L/(\rho t^2 \theta_0^2 c^5)$.  For a relativistic jet head in a medium with $\rho = A/r^2$, this is $\tilde L = L/(A \theta_0^2 c^3)$.

Thus, the condition can be expressed in terms of the quantity $\lambda$:

\begin{equation}
    \lambda \equiv \frac{L}{A \theta_0^2 c^3}.
\end{equation}

If $\lambda < \lambda_{\rm crit}$, the jet will be choked as soon as the engine is shut off.  \cite{2011ApJ...740..100B} predict that this $\lambda_{\rm crit} \sim \theta_0^{-4/3} \approx 10$, but we will determine its value for a single choice of $\theta_0 = 0.16$.  In reality, $\lambda_{\rm crit}$ should depend on $\theta_0$.

\section{Numerical Calculations} \label{sec:numerics}

For $\lambda$ below some critical value, the jet will be choked when the engine shuts off.  In this study, we calculate this critical value $\lambda_{\rm crit}$ via a direct numerical calculation of a jet breaking out of a star.

Numerical calculations are carried out using the JET code \citep{2011ApJS..197...15D, 2013ApJ...775...87D}. The JET code solves the equations of relativistic gas dynamics using a moving mesh.  The mesh motion allows for evolution of very high Lorentz factors over many orders of magnitude of expansion.

\subsection{Initial Conditions}

The initial conditions assume a ``generic progenitor" model, identical to the stellar model used in \cite{2018ApJ...860...38B}.  Specifically, the initial density is given by 

\begin{equation}
\rho_{\rm init}(r) = \left\{ \begin{array}
				{l@{\quad \quad}l}
				\rho_{\rm star}(r) + A/R_0^2 & r < R_0	\\  
    			A/r^2 & r > R_0 \\ 
    			\end{array} \right.
\end{equation}
where

\begin{equation}
\rho_{\rm star}(r) = \frac{0.0615 M_{\rm prog}}{R_0^3} (R_0/r)^{2.65}(1-r/R_0)^{3.5},
\end{equation}
$M_{\rm prog}$ is the progenitor mass after core collapse, and $R_0$ is the stellar radius.  Fiducial values for these quantities are $M_{\rm prog} = 2.5 M_{\sun}$ and $R_0 = 1.6 R_{\sun}$, but because of hydrodynamical scale invariance, these values do not need to be explicitly chosen.  Just as in \cite{2018ApJ...860...38B}, material interior to $r < 1.5 \times 10^{-3} R_0$ is reduced significantly in density, assuming the core has collapsed and left a cavity behind.
 
 Pressure is chosen to be initially negligible and velocity is set to zero everywhere.  Note the CSM extends to infinity, and not to some finite ``extended stellar radius".  Therefore, if $A$ is sufficiently large, the jet will be choked when the engine has been turned off, regardless of how far the jet has propagated.
 
The engine is injected as a source term, as in \cite{2015ApJ...806..205D, 2018ApJ...860...38B, 2018ApJ...866....3D}.  In this study, however, the engine luminosity is a constant until the time $t = T$, at which point the engine is instantaneously shut off.  This is to ensure that the engine shut-off occurs at a well-defined instant in time (although real GRB engines probably do not behave in this way).  For our calculations, the jet power is $L = 10^{-4} M_0 c^3 / R_0$, or $1.2 \times 10^{50}$ erg/s.  The opening angle of the engine is $\theta_0 = 0.16$, but the head of the jet may become narrower than this.

We perform several different calculations.  First, a test run is performed where the engine is shut off before breakout from the stellar surface ($T = 0.25 R_0/c = 0.93$ seconds), to demonstrate that the star can indeed choke the jet if the engine is shut off too soon.

Second, a suite of calculations is performed where the engine is shut off significantly after breakout from the stellar surface ($T = 3 R_0/c = 11$ seconds), but where the external density is varied to determine how dense a CSM is necessary to choke the jet.  The choices of wind parameter range from $A = 10^{-9} M_0/R_0 = 4 \times 10^{13}$ g/cm to $10^{-2} M_0/R_0 = 4 \times 10^{20}$ g/cm, with eight total choices of $A$, set a factor of ten apart.  Our calculation with $A = 10^{-6} M_0/R_0$ is equivalent to $0.01 M_{\sun}$ of CSM spread out over $4 \times 10^{13}$ cm.

Third, for our set of calculations with $A = 10^{-6} M_0/R_0$ (the highest observed densities), we run a handful of models with the same energy but wider injection angles ($\theta_0 = 0.16, 0.2, 0.22,$ and $0.24$), to see whether any of these jets choke or appreciably spread in the CSM.

For these tests, the total energy in the ``early shut-off" model is $1.2 \times 10^{50}$ ergs, and the total energy deposited in each of the other models is $1.4 \times 10^{51}$ erg.

\begin{figure}
\epsscale{1.15}
\plotone{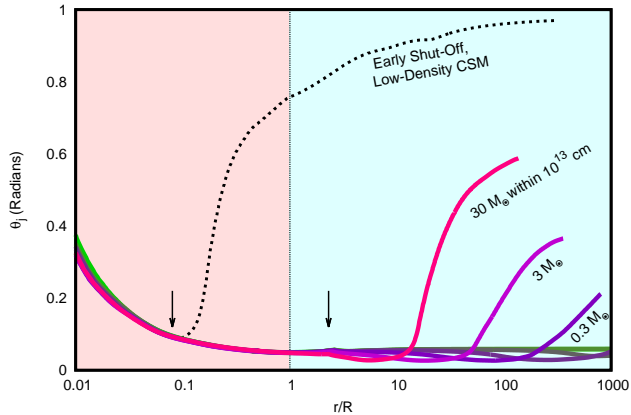}
\caption{ Evolution of opening angle $\theta_j$ of the jet with forward shock radius $r$ (scaled by stellar radius $R$), for the eight different CSM densities considered.  The dashed curve is the ``Early Shut-Off" model whose engine is shut off before the jet has escaped the star.  All other models have engines that are shut off after breakout from the stellar surface (shut-off times indicated by black arrows); in that case, one requires a very high-density CSM ($A > 4 \times 10^{19}$ g/cm or $\lambda < 10$) in order to choke the jet.
\label{fig:angles} }
\end{figure}

\subsection{Diagnostics}

We measure the opening angle of the jet with time, calculated from the isotropic equivalent energy, as in \cite{2018ApJ...865...94D}.  The solid angle $\Omega$ subtended by the jet is computed by

\begin{equation}
\Omega = 4 \pi E / E_{\rm iso},
\end{equation}
which, given the following expression to compute $E_{\rm iso}$,

\begin{equation}
E_{\rm iso} = 4 \pi \frac{ \int (dE/d\Omega)^2 d\Omega }{ \int (dE/d\Omega) d\Omega }
\end{equation}
results in the following formula for $\theta_j$, which is evaluated at regular intervals:

\begin{equation}
{\rm sin}(\theta_j/2) = \frac{ E }{ \sqrt {4 \pi \int (dE/d\Omega)^2 d\Omega } }.
\end{equation}

The opening angle $\theta_j$ is measured as a function of time, to determine how much the jet widens after the engine is turned off.

\section{Results and Discussion} \label{sec:disc}

\begin{figure}
\epsscale{1.15}
\plotone{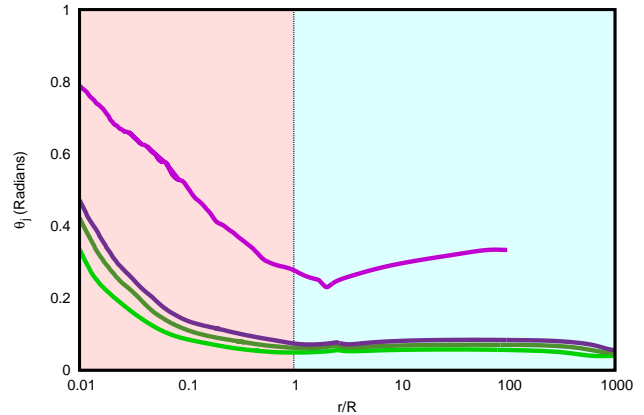}
\caption{ Choosing a high CSM density ($A = 4 \times 10^{16}$ g/cm or $0.02 M_{\sun}$ within $10^{14}$ cm) and fixed GRB energy $E = 1.4 \times 10^{51}$ ergs but varying the injected opening angle ($\theta_0 = 0.16, 0.2, 0.22,$ and $0.24$; the final opening angle is a nonlinear function of the injection angle).  All models that successfully make it out of the star do not choke or spread significantly in the CSM.
\label{fig:angles2} }
\end{figure}

Figure \ref{fig:angles} shows opening angle as a function of forward shock radius for our suite of jet models.  The ``early shut-off" jet (dashed curve) begins to spread immediately after the engine is turned off, as the jet is still propagating through the star at the shut-off time; this jet is properly choked.

For the other models, as the engine is turned off after the shock has expanded to $r>R_0$, the jet spreading depends on the density.  In particular, none of the models experienced significant spreading of the jet, except for the three highest-density calculations ($A = 10^{-4} M_0/R_0$, $10^{-3} M_0/R_0$, and $10^{-2} M_0/R_0$).  In fact, as we discuss in Section 1.4, for all but the highest-density case this spreading does not appear to be due to jet ``choking", but simply the jet spreading slowly as it decelerates, like a standard afterglow jet.  Only the highest-density CSM exhibited enough spreading to call the jet ``choked".  The critical density appears to be $A = 10^{-3} M_0/R_0 = 4 \times 10^{19}$ g/cm, which corresponds to $\lambda = \lambda_{\rm crit} = 10$.  This is consistent with the prediction of \cite{2011ApJ...740..100B} for the transition between a collimated vs. uncollimated jet.  Such a density necessitates several solar masses of CSM within $10^{13}$ cm of the progenitor, which is several orders of magnitude higher than any CSM density observed.

When running the suite of calculations varying the opening angle with fixed density and energy consistent with observed values (Figure \ref{fig:angles2}), essentially no choking or spreading was observed for any model that was not choked before escaping the star.

\begin{figure}
\epsscale{1.15}
\plotone{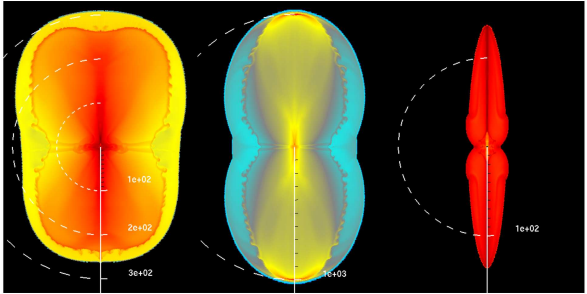}
\caption{ Energy density in several different models, long after breakout from the star, and long after the engine is shut off; $t = 10^{3} R_0/c \approx 1$ hour.  All models have been interacting with the CSM for the entire hour after break-out.  Left: The ``Early Shut-Off" model, properly choked before escaping the star.  The outflow is more spherical and sub-relativistic; the forward shock is only propagating at about $0.25 c$.  Center: A low-density CSM model ($A = 10^{-9} M_0/R_0$) whose engine was shut off after break-out; the CSM density was not high enough to choke the jet.  The jet is narrow and relativistic ($\Gamma = 285$).  Right: A jet which was somewhat choked by the CSM ($v = 0.25 c$).  The jet core is still collimated but a large portion of the energy is more spherically distributed, at low velocities.  This required a very high density; $A = 10^{-2} M_0/R_0 \approx 4 \times 10^{20}$ g/cm.  
\label{fig:images} }
\end{figure}

\subsection{``Choked" vs. Spreading}
While jets do not appear to ever be ``choked" in the CSM for observed GRB energies, non-choked jets can still have their opening angle widened simply due to the spreading of the jet as it decelerates.  In this case, the criterion for jet spreading has nothing to do with the duration of the engine.  It only has to do with the Lorentz factor of the jet after entraining the CSM, and how this Lorentz factor compares to $1/\theta_j$.  The Lorentz factor is given by

\begin{equation}
    \Gamma \approx 3 \sqrt{
    \frac{E_{\rm iso}}{M_{\rm CSM} c^2}}
\end{equation}
\citep{1976PhFl...19.1130B}.

For a CSM mass of $0.01 M_{\sun}$, the final Lorentz factor can range from $1$ to $50$, depending on $E_{\rm iso}$.  Jets that attain low Lorentz factors will still require a few orders of magnitude of expansion to spread \citep{2009ApJ...698.1261Z, 2018ApJ...865...94D}, so this low Lorentz factor must be attained early in order to significantly change the jet opening angle.  Typically, non-choked jets stay reasonably narrow ($\theta_j < 20$ degrees) until they decelerate to non-relativistic velocities \citep[see Figure 1 of][]{2018ApJ...865...94D}.  Thus, only the very weakest GRB jets with $E_{\rm iso} \sim 10^{51}$ ergs could widen significantly via spreading in the CSM, independent of the duration of the engine.

For reference, the run that exhibited spreading to $0.4$ radians in Figure \ref{fig:angles} has $E_{\rm iso}/(M_{\rm CSM} c^2) = 0.03$ (where $E_{\rm iso}$ was computed before spreading), comparable to a very low-energy GRB with $E_{\rm iso} = 10^{51}$ erg in $0.01 M_{\sun}$ of CSM.  So, if spreading  occurs it will only be for the lowest-energy jets in the highest-density CSM, and even in this case the spreading is quite modest, as the final opening angle is about 20 degrees.

\subsection{Possible Connection to Afterglow Plateaus}

Even if the jet is not choked, it may entrain a significant amount of mass, decelerating it to a lower Lorentz factor than it emerged with from the star.  For example, a typical jet will sweep up $1\%$ of the surrounding CSM, which for a $10^{51}$ erg jet in a $0.01 M_{\sun}$ CSM, results in a jet Lorentz factor of $\Gamma \sim 10$ upon emerging from the CSM.  This jet could still remain somewhat collimated until it swept up significant mass during the afterglow phase.  This would give rise to a long plateau as the jet coasts at a modest Lorentz factor, similar to the ``top-heavy" plateau model of \cite{2014ApJ...791L...1D}.  According to that study, the deceleration time (according to the observer) is sensitive to the Lorentz factor:

\begin{equation}
    t^{\rm obs}_{\rm decel} \sim \frac{E_{\rm iso}}{\rho_{\rm ext} r^2 c^3 \Gamma^4} \sim \frac{10^{10} \rm s}{\Gamma^4},
\end{equation}
where $\rho_{\rm ext}$ is the (presumably much lower) CSM density at larger radii.

If the jet emerges from the dense component of the CSM with a Lorentz factor $\Gamma \approx 10$, this could lead to a plateau duration of up to $10^{6}$ seconds, as long as several days.  In practice, many CSM environments may not be this dense, only decelerating the jet to a Lorentz factor of $30$ or so, leading to more modest plateau durations.

One way of confirming this scenario would be to make observations of the reverse shock during the afterglow, to measure the Lorentz factor of the jet during the plateau \citep{2013ApJ...776..119L, 2017ApJ...848...69A}.  So far, the handful of direct observations of reverse shocks have all coincided with afterglows that did not exhibit an X-ray plateau.  Similarly, one would expect (for the same CSM mass) that the plateau duration should be inversely proportional to $E_{\rm iso}$:

\begin{equation}
    t^{\rm obs}_{\rm decel} \propto \frac{M_{\rm CSM}^2 c}{\rho_{\rm ext} r^2 E_{\rm iso}}.
\end{equation}

This could be a partial explanation for the observed Dainotti relation \citep{2008MNRAS.391L..79D} where the luminosity of the afterglow at the end of the plateau correlates inversely with the plateau duration.

\subsection{Implications for the Prompt Emission}

A `non-choked' jet might still be unobservable if it is sufficiently decelerated before gamma rays can be emitted.  In this case, it is possible that a very dense CSM might hide the burst sufficiently well that it effectively doesn't matter whether the jet was choked.  Such a jet would still produce an afterglow, but without the prompt emission it would be much harder to detect (but see \citealt{Berger2013,Cenko2013,Cenko2015,Ho2018}).  Alternatively, it is possible that this still produces a low luminosity GRB via shock breakout, similar to what was  predicted by \cite{2015ApJ...807..172N}, but in contrast with that picture, the shock energy would not be redistributed spherically; it would still be concentrated in the direction of the jet.

However, for a modestly dense CSM (say, a few times $10^{-3} M_{\sun}$) or a high-energy jet ($E_{\rm iso} \sim 10^{54}$ ergs), it might be possible to produce a soft gamma ray burst (or X-ray burst) coupled with an afterglow exhibiting a long plateau.
This is very similar to the ``dirty fireball'' model of \cite{Dermer2000}.
However, in that model, the jet is thought to entrain material within the star itself.

A potentially observable signature would be a correlation between soft gamma ray bursts and longer plateaus in the afterglow (and if a reverse shock is observed, a modest Lorentz factor for the jet).  If this dense CSM were responsible for most X-ray plateaus, one would expect afterglows with plateaus to exhibit lower $E_{\rm iso}$ and softer prompt emission than afterglows without plateaus.

~

\acknowledgments

We thank Tanmoy Laskar, Kate Alexander, Edo Berger and Andrew MacFadyen for helpful comments and discussions.  We especially thank Ehud Nakar for very detailed comments and suggestions for additional calculations that have helped to bolster our arguments.

High-resolution calculations were provided by the NASA High-End Computing (HEC) Program through the NASA Advanced Supercomputing (NAS) Division at Ames Research Center.  

A.Y.Q.H. is supported by a National Science Foundation Graduate Research Fellowship under Grant No.\,DGE‐1144469.
This work was supported by the GROWTH project funded by the National Science Foundation under PIRE Grant No.\,1545949.

\bibliographystyle{apj} 
\bibliography{ms}

\begin{thebibliography}{}
\expandafter\ifx\csname natexlab\endcsname\relax\def\natexlab#1{#1}\fi

\bibitem[{{Alexander} {et~al.}(2017){Alexander}, {Laskar}, {Berger},
  {Guidorzi}, {Dichiara}, {Fong}, {Gomboc}, {Kobayashi}, {Kopac}, \&
  {Mundell}}]{2017ApJ...848...69A}
{Alexander}, K.~D., {Laskar}, T., {Berger}, E., {et~al.} 2017, \apj, 848, 69

\bibitem[{{Barnes} {et~al.}(2018){Barnes}, {Duffell}, {Liu}, {Modjaz},
  {Bianco}, {Kasen}, \& {MacFadyen}}]{2018ApJ...860...38B}
{Barnes}, J., {Duffell}, P.~C., {Liu}, Y., {et~al.} 2018, \apj, 860, 38

\bibitem[{{Berger} {et~al.}(2013){Berger}, {Leibler}, {Chornock}, {Rest},
  {Foley}, {Soderberg}, {Price}, {Burgett}, {Chambers}, \&
  {Flewelling}}]{Berger2013}
{Berger}, E., {Leibler}, C.~N., {Chornock}, R., {et~al.} 2013, \apj, 779, 18

\bibitem[{{Blandford} \& {McKee}(1976)}]{1976PhFl...19.1130B}
{Blandford}, R.~D., \& {McKee}, C.~F. 1976, Physics of Fluids, 19, 1130

\bibitem[{{Bromberg} {et~al.}(2011){Bromberg}, {Nakar}, {Piran}, \&
  {Sari}}]{2011ApJ...740..100B}
{Bromberg}, O., {Nakar}, E., {Piran}, T., \& {Sari}, R. 2011, \apj, 740, 100

\bibitem[{{Campana} {et~al.}(2006){Campana}, {Mangano}, {Blustin}, {Brown},
  {Burrows}, {Chincarini}, {Cummings}, {Cusumano}, {Della Valle}, {Malesani},
  {M{\'e}sz{\'a}ros}, {Nousek}, {Page}, {Sakamoto}, {Waxman}, {Zhang}, {Dai},
  {Gehrels}, {Immler}, {Marshall}, {Mason}, {Moretti}, {O'Brien}, {Osborne},
  {Page}, {Romano}, {Roming}, {Tagliaferri}, {Cominsky}, {Giommi}, {Godet},
  {Kennea}, {Krimm}, {Angelini}, {Barthelmy}, {Boyd}, {Palmer}, {Wells}, \&
  {White}}]{2006Natur.442.1008C}
{Campana}, S., {Mangano}, V., {Blustin}, A.~J., {et~al.} 2006, \nat, 442, 1008

\bibitem[{{Cano} {et~al.}(2017){Cano}, {Wang}, {Dai}, \&
  {Wu}}]{2017AdAst2017E...5C}
{Cano}, Z., {Wang}, S.-Q., {Dai}, Z.-G., \& {Wu}, X.-F. 2017, Advances in
  Astronomy, 2017, 8929054

\bibitem[{{Cenko} {et~al.}(2013){Cenko}, {Kulkarni}, {Horesh}, {Corsi}, {Fox},
  {Carpenter}, {Frail}, {Nugent}, {Perley}, \& {Gruber}}]{Cenko2013}
{Cenko}, S.~B., {Kulkarni}, S.~R., {Horesh}, A., {et~al.} 2013, \apj, 769, 130

\bibitem[{{Cenko} {et~al.}(2015){Cenko}, {Urban}, {Perley}, {Horesh}, {Corsi},
  {Fox}, {Cao}, {Kasliwal}, {Lien}, \& {Arcavi}}]{Cenko2015}
{Cenko}, S.~B., {Urban}, A.~L., {Perley}, D.~A., {et~al.} 2015, \apj, 803, L24

\bibitem[{{Corsi} {et~al.}(2014){Corsi}, {Ofek}, {Gal-Yam}, {Frail},
  {Kulkarni}, {Fox}, {Kasliwal}, {Sullivan}, {Horesh}, {Carpenter}, {Maguire},
  {Arcavi}, {Cenko}, {Cao}, {Mooley}, {Pan}, {Sesar}, {Sternberg}, {Xu},
  {Bersier}, {James}, {Bloom}, \& {Nugent}}]{2014ApJ...782...42C}
{Corsi}, A., {Ofek}, E.~O., {Gal-Yam}, A., {et~al.} 2014, \apj, 782, 42

\bibitem[{{Dainotti} {et~al.}(2008){Dainotti}, {Cardone}, \&
  {Capozziello}}]{2008MNRAS.391L..79D}
{Dainotti}, M.~G., {Cardone}, V.~F., \& {Capozziello}, S. 2008, \mnras, 391,
  L79

\bibitem[{{Dermer} {et~al.}(2000){Dermer}, {Chiang}, \& {Mitman}}]{Dermer2000}
{Dermer}, C.~D., {Chiang}, J., \& {Mitman}, K.~E. 2000, \apj, 537, 785

\bibitem[{{Duffell} \& {Laskar}(2018)}]{2018ApJ...865...94D}
{Duffell}, P.~C., \& {Laskar}, T. 2018, \apj, 865, 94

\bibitem[{{Duffell} \& {MacFadyen}(2011)}]{2011ApJS..197...15D}
{Duffell}, P.~C., \& {MacFadyen}, A.~I. 2011, \apjs, 197, 15

\bibitem[{{Duffell} \& {MacFadyen}(2013)}]{2013ApJ...775...87D}
---. 2013, \apj, 775, 87

\bibitem[{{Duffell} \& {MacFadyen}(2014)}]{2014ApJ...791L...1D}
---. 2014, \apjl, 791, L1

\bibitem[{{Duffell} \& {MacFadyen}(2015)}]{2015ApJ...806..205D}
---. 2015, \apj, 806, 205

\bibitem[{{Duffell} {et~al.}(2018){Duffell}, {Quataert}, {Kasen}, \&
  {Klion}}]{2018ApJ...866....3D}
{Duffell}, P.~C., {Quataert}, E., {Kasen}, D., \& {Klion}, H. 2018, \apj, 866,
  3

\bibitem[{{Filippenko}(1997)}]{Filippenko1997}
{Filippenko}, A.~V. 1997, \araa, 35, 309

\bibitem[{{Ho} {et~al.}(2018){Ho}, {Kulkarni}, {Nugent}, {Zhao}, {Rusu},
  {Cenko}, {Ravi}, {Kasliwal}, {Perley}, \& {Adams}}]{Ho2018}
{Ho}, A. Y.~Q., {Kulkarni}, S.~R., {Nugent}, P.~E., {et~al.} 2018, \apj, 854,
  L13

\bibitem[{{Ho} {et~al.}(2019){Ho}, {Goldstein}, {Schulze}, {Khatami}, {Perley},
  {Ergon}, {Gal-Yam}, {Corsi}, {Andreoni}, {Barbarino}, {Bellm},
  {Blagorodnova}, {Bright}, {Burns}, {Cenko}, {Cunningham}, {De}, {Dekany},
  {Dugas}, {Fender}, {Fransson}, {Fremling}, {Goldstein}, {Graham}, {Hale},
  {Horesh}, {Hung}, {Kasliwal}, {Kuin}, {Kulkarni}, {Kupfer}, {Lunnan},
  {Masci}, {Ngeow}, {Nugent}, {Ofek}, {Patterson}, {Petitpas}, {Rusholme},
  {Sai}, {Sfaradi}, {Shupe}, {Sollerman}, {Soumagnac}, {Tachibana}, {Taddia},
  {Walters}, {Wang}, {Yao}, \& {Zhang}}]{2019arXiv190411009H}
{Ho}, A.~Y.~Q., {Goldstein}, D.~A., {Schulze}, S., {et~al.} 2019, arXiv
  e-prints, arXiv:1904.11009

\bibitem[{{Izzo} {et~al.}(2019){Izzo}, {de Ugarte Postigo}, {Maeda},
  {Th{\"o}ne}, {Kann}, {Della Valle}, {Sagues Carracedo}, {Micha{\l}owski},
  {Schady}, {Schmidl}, {Selsing}, {Starling}, {Suzuki}, {Bensch}, {Bolmer},
  {Campana}, {Cano}, {Covino}, {Fynbo}, {Hartmann}, {Heintz}, {Hjorth},
  {Japelj}, {Kami{\'n}ski}, {Kaper}, {Kouveliotou}, {Kru{\.Z}y{\'n}ski},
  {Kwiatkowski}, {Leloudas}, {Levan}, {Malesani}, {Micha{\l}owski},
  {Piranomonte}, {Pugliese}, {Rossi}, {S{\'a}nchez-Ram{\'\i}rez}, {Schulze},
  {Steeghs}, {Tanvir}, {Ulaczyk}, {Vergani}, \&
  {Wiersema}}]{2019Natur.565..324I}
{Izzo}, L., {de Ugarte Postigo}, A., {Maeda}, K., {et~al.} 2019, \nat, 565, 324

\bibitem[{{Laskar} {et~al.}(2013){Laskar}, {Berger}, {Zauderer}, {Margutti},
  {Soderberg}, {Chakraborti}, {Lunnan}, {Chornock}, {Chandra}, \&
  {Ray}}]{2013ApJ...776..119L}
{Laskar}, T., {Berger}, E., {Zauderer}, B.~A., {et~al.} 2013, \apj, 776, 119

\bibitem[{{MacFadyen} {et~al.}(2001){MacFadyen}, {Woosley}, \&
  {Heger}}]{2001ApJ...550..410M}
{MacFadyen}, A.~I., {Woosley}, S.~E., \& {Heger}, A. 2001, \apj, 550, 410

\bibitem[{{Margutti} {et~al.}(2015){Margutti}, {Guidorzi}, {Lazzati},
  {Milisavljevic}, {Kamble}, {Laskar}, {Parrent}, {Gehrels}, \&
  {Soderberg}}]{2015ApJ...805..159M}
{Margutti}, R., {Guidorzi}, C., {Lazzati}, D., {et~al.} 2015, \apj, 805, 159

\bibitem[{{M{\'e}sz{\'a}ros} \& {Waxman}(2001)}]{Meszaros2001}
{M{\'e}sz{\'a}ros}, P., \& {Waxman}, E. 2001, Physical Review Letters, 87,
  171102

\bibitem[{{Modjaz} {et~al.}(2016){Modjaz}, {Liu}, {Bianco}, \&
  {Graur}}]{Modjaz2016}
{Modjaz}, M., {Liu}, Y.~Q., {Bianco}, F.~B., \& {Graur}, O. 2016, \apj, 832,
  108

\bibitem[{{Nakar}(2015)}]{2015ApJ...807..172N}
{Nakar}, E. 2015, \apj, 807, 172

\bibitem[{{Nakar}(2019)}]{2019Natur.565..300N}
---. 2019, \nat, 565, 300

\bibitem[{{Nakar} \& {Piro}(2014)}]{2014ApJ...788..193N}
{Nakar}, E., \& {Piro}, A.~L. 2014, \apj, 788, 193

\bibitem[{{Piro}(2015)}]{Piro2015}
{Piro}, A.~L. 2015, \apj, 808, L51

\bibitem[{{Salas} {et~al.}(2013){Salas}, {Bauer}, {Stockdale}, \&
  {Prieto}}]{2013MNRAS.428.1207S}
{Salas}, P., {Bauer}, F.~E., {Stockdale}, C., \& {Prieto}, J.~L. 2013, \mnras,
  428, 1207

\bibitem[{{Sobacchi} {et~al.}(2017){Sobacchi}, {Granot}, {Bromberg}, \&
  {Sormani}}]{Sobacchi2017}
{Sobacchi}, E., {Granot}, J., {Bromberg}, O., \& {Sormani}, M.~C. 2017, \mnras,
  472, 616

\bibitem[{{Soderberg} {et~al.}(2006){Soderberg}, {Chevalier}, {Kulkarni}, \&
  {Frail}}]{2006ApJ...651.1005S}
{Soderberg}, A.~M., {Chevalier}, R.~A., {Kulkarni}, S.~R., \& {Frail}, D.~A.
  2006, \apj, 651, 1005

\bibitem[{{Terreran} {et~al.}(2019){Terreran}, {Margutti}, {Bersier},
  {Brimacombe}, {Caprioli}, {Challis}, {Chornock}, {Coppejans}, {Dong},
  {Guidorzi}, {Hurley}, {Kirshner}, {Migliori}, {Milisavljevic}, {Palmer},
  {Prieto}, {Tomasella}, {Marchant}, {Pastorello}, {Shappee}, {Stanek},
  {Stritzinger}, {Benetti}, {Demarchi}, {Elias-rosa}, {Gall}, {Harmanen}, \&
  {Mattila}}]{2019arXiv190502226T}
{Terreran}, G., {Margutti}, R., {Bersier}, D., {et~al.} 2019, arXiv e-prints,
  arXiv:1905.02226

\bibitem[{{Whitesides} {et~al.}(2017){Whitesides}, {Lunnan}, {Kasliwal},
  {Perley}, {Corsi}, {Cenko}, {Blagorodnova}, {Cao}, {Cook}, {Doran},
  {Frederiks}, {Fremling}, {Hurley}, {Karamehmetoglu}, {Kulkarni}, {Leloudas},
  {Masci}, {Nugent}, {Ritter}, {Rubin}, {Savchenko}, {Sollerman}, {Svinkin},
  {Taddia}, {Vreeswijk}, \& {Wozniak}}]{2017ApJ...851..107W}
{Whitesides}, L., {Lunnan}, R., {Kasliwal}, M.~M., {et~al.} 2017, \apj, 851,
  107

\bibitem[{{Woosley} \& {Bloom}(2006)}]{WoosleyBloom}
{Woosley}, S.~E., \& {Bloom}, J.~S. 2006, \araa, 44, 507

\bibitem[{{Zhang} \& {MacFadyen}(2009)}]{2009ApJ...698.1261Z}
{Zhang}, W., \& {MacFadyen}, A. 2009, \apj, 698, 1261

\end{thebibliography}

%\begin{comment}
%\end{comment}

\end{document}